# New conceptions on the light wave theory in a moving frame of reference.

**Daniel Lima Nascimento**

Institute of Physics, University of Brasilia, Brazil, daniel@fis.unb.br

**Abstract:** A reanalysis of the effect of a moving frame of reference on the Maxwell's equations of light is done, showing that the null result of the Michelson-Morley's experiment of 1887 on the possibility of detecting the variation of the velocity of light with respect to a moving observer, i.e., therefore with no need of new *ad hoc* hypotheses to explain it in the classical framework. In fact, it will be shown that behind those equations is hidden the fascinating phenomenon of the absolute motion. (This is an old paper which I did not succeed in publishing 20 years ago and since then has been dismissed, but I hope somebody else can have some fun in reading it.)

## 1. Introduction.

As is known, the apparently null result of the Michelson-Morley's interference experiment (MMIE) performed in 1887 [1] put in check the classically fundamental notion of absolute motion. The detection of it through the variation of the velocity of light with respect to the motion of the Earth had been proposed by Lord Kelvin [2] because the so called Galilean Relativity – the invariance of the mechanical laws with respect to uniform translations of the coordinates – no mechanical system could be used for such a detection. Yet, with respect to electromagnetic systems the things could be completely different, because Maxwell himself believed that the usual form of his equations would be valid only in a frame of reference which stayed at absolute rest [3]. Since, of course, the Earth's frame could not be this frame, it would be possible in principle to detect its motion trough absolute space through the variation of wave properties of light, including the ordinary Doppler effect, when measured at different directions with regard to the direction of the Earth's motion – as it was attempted in the MMIE mentioned above.

As a consequence, after the initial shock caused by the non detection of the effect (whose size was too small, of order $(v_{Earth}/c)^2 \sim 10^{-8}$), some miraculous *ad hoc* hypotheses were proposed in order to try to save the classical theory. For example, the Ether drift together with the Earth, which was proposed by Michelson himself [4], or the mysterious contraction of the interferometer arm that were in the direction of motion of the Earth, proposed by Fitzgerald and Lorentz independently [5]. Against this later, an objection was immediately risen by Poincaré [6] that it had been necessary a new hypothesis for taking theoretically into account a new phenomenon, instead of having it been deduced from a general *a priori* principle – the Maxwell equations (ME) in this case. And this theoretical violation would represent a violation of the whole schema of explanation of the known phenomena of the nature by the classical picture of the physical reality.

With the aim of surmount this objection, after several unsuccessful attempts, Lorentz published in 1904 [5] a long paper in which he practically elucidated the problem, by using the transformations of coordinates that later would receive his name. However, and unfortunately for classical physics, he made a mistaken in the definition of the transformation of the electron charge density, which led him to wrong velocity transformations between the rest and moving frames of reference. Further, he gave a misleading emphasis to dynamic features of the electron motion instead to wave features of light – which were in fact the most relevant properties. Therefore, his analysis could not give a satisfactory explanation of the interference problem – and this was regrettable because the interference problem contained the roots of the solution of the problem of the absolute motion without making use of any ad hypothesis. An attempt of getting such an *a priori* formulation is the main concern of this paper.

But the things at the beginning of the XX century followed in fact the way of the strong change of direction proposed by the neopositivists, who followed the subjective inclinations of E. Mach [7]. In this direction, A. Einstein [8] published in 1905 his solution of the problem of the electromagnetic phenomena in a moving frame, which had the merit of answering all the theoretical objections, in such a way that Lorentz himself considered it as a definite theory. However, the Einstein's operationalistic theory [9] which, epistemologically speaking, started from the formula "reality=appearance"[1] which, although useful, had only the effect of hiding the true problem – namely of the existence and detection of the absolute motion – because he took as definitive the negative result of the MMIE. Moreover, he postulated [10], that is, made a new *ad hoc* , hypothesis, that the velocity of light were not affected by the motion of the observer so that it could not be detected any effect of time delay among light waves that followed different directions of motion within the interferometer with respect to the direction of the Earth's motion. On the other hand, he postulated also the more reasonable principle of general invariance of the physical laws in inertial frames, from which he succeeded in deducing the Lorentz transformations.

However, the neopositivistic Einstein's solution, even intending to remove metaphysical conceptions – as absolute space, time and motion – ended by introducing even more, through his conceptions of relative space, time and motion, which, besides being equally metaphysical, are in addition subjective because they depend upon the point of view of a particular observer. In fact it is impossible to avoid metaphysics from physics completely because its starting point is the belief that Nature may be explained or described by general mathematical laws and, as clearly was shown by D. Hume [11], there is no logical way of proving such an assumption. In order to overcome this *a priori* impossibility of obtaining an absolute knowledge, we must search for a mathematical representation of Natures which contains the least of metaphysics and the most of objectiveness, a formula which is the aim of Classical Physics [12].

Its is also the aim of this work trying to search for the lost realism hidden behind the current neopositivistic formulation of physics. Thus, it will be constructed a formulation in which the problem of the absolute motion and its relationship with the wave properties of light *may be deduced rigorously from the ME*, following a line of thought similar to that followed by Lorentz in 1904, *without any ad hoc hypothesis*.

First of all, it will be shown that the null result of the MMIE may be deduced rightly from the ME and therefore it is not a surprising result, at least mathematically speaking, it being in fact due to an automatic compensation of the propagation times of light waves when propagating through an optically homogeneous medium, independently of the direction that may do with the velocity of the Earth. Such a compensation will be seen to be due to the form of composition between the velocity of a light wave and that of the Earth, which may not be the ordinary vector rule of composition of velocities – as was wrongly assumed by Lorentz, Michelson and others, which was the true source of misunderstanding and confusion with regard to the analysis of the MMIE – but instead of this the composition is done through a specific electromagnetic rule, derived directly from the analysis of the ME when undergone to a uniform translation of the coordinates.

---

[1] Such an Aristotelian formula was condemned even in the Greek Antiquity as representing a naïve thought and this adoption by the Catholic Roman Church led to the anti-scientific minded Middle Age [11].

Secondly, it will be briefly commented that the use of "ideal" clocks instead real ones in the Einstein's imaginary experiments that founded his "relativistics" conceptions, is the responsible by the metaphysical and subjective features of the Special Theory of Relativity (STR). The dilation time in this theory it will be seen to appear only in electromagnetical clocks, but never in mechanical clocks. Therefore, the "delay" of time in STR is a completely physical effect and not a metaphysical one as suggested in Einstein's 1905 analysis. Hence, an accurate measure of the universal or absolute time can be furnished either by a good mechanical clock or rather by a suitably motion-compensated electromagnetic clock.

At the final of this work we shall hit the conclusion that STR is nothing but a disguised form of a Theory of Absolute Motion, it being for example the so called "Twins Paradox" [8] a paradox only in the neopositivistic schema. On the contrary, in a realistic picture this paradox would be a strong evidence of the asymmetry between "rest" and "moving" situations. Other empirical evidences would be the very MMIE and other delay experiments like those of particle decays and atomic clocks, in particular the 1938 Yves and Stillwell experiments [13].

To close the paper it is proposed a possible "positive" experiment for the detection of the motion of the Earth through the properties of light waves by using a modified type interferometer, which can avoid the natural compensation of the optical paths in a homogeneous medium, which was responsible by the null effect in the original MMIE. This experiment may contribute to bring back the beauty and objectiveness that Physics had at the end of the XIX century [14,15].

## 2. Transformation of the Maxwell's and wave equations under a Galileo's transformation of coordinates.

As early as in 1860 [3] – when Maxwell published his system of equations that unified in an only theoretical-mathematical body the description or explanation of magnetism and electricity, which were then considered as independent phenomena – questions arose concerning to which frame of reference the Maxwell's equations were written. Maxwell himself had not any doubt in answering that they had the specific form then published only in a coordinate system associated to a frame of reference which stayed at absolute rest and that, of course, the Earth's frame could not be that frame. Moreover, the Maxwell's equations would have to present modifications due the existence of the known motions of the Earth, in particular the translation motion, that could be eventually be detected through the variations of electromagnetic wave properties, that is, the light waves, because light had then just been recognized as a kind of electromagnetic wave. By this time, however, most of the science researchers had the belief that light was an elastic vibration, like the sound waves, and following this belief they postulated the existence of an interstellar medium, the so called luminous ether, through which the electromagnetic vibrations could propagate. This assumption however was responsible by serious difficulties and misunderstandings in the analisys of the light properties in moving frames. Today we know that an electromagnetic wave is a self-oscillating arrangement of mutually induced electric and magnetic fields which do not need any physical medium to exist and propagate, so that it can propagate even in the vacuum or empty space. According to this, we may call from now on the own empty space as being the absolute space – which was postulated by Newton in his Principia Mathematica [16] – it being immovable, homogeneous and completely independent of any physical process that occurs *in it*.

Following this line of reasoning we suppose that only in an absolute rest frame (ARF), the ME for light phenomena could be written in the standard form [17] below

$$\nabla \times \mathbf{E} = -\tfrac{1}{c}\frac{\partial \mathbf{B}}{\partial t}, \qquad \nabla \cdot \mathbf{E} = 0, \qquad (1a,1b)$$

$$\nabla \times \mathbf{B} = \tfrac{1}{c}\frac{\partial \mathbf{E}}{\partial t}, \qquad \nabla \cdot \mathbf{B} = 0, \qquad (1c,1d).$$

From these equations it can easily be deduced the differential equation for the electric field vector $\mathbf{E}$, by taking the curl of (1a) and using (1c), that is

$$\Delta \mathbf{E} - \tfrac{1}{c^2}\frac{\partial^2 \mathbf{E}}{\partial t^2} = 0, \qquad (2)$$

in which the Laplacian operator $\Delta$ is written in the absolute coordinates (x,y,z) of the absolute frame of reference S and t is the absolute time, namely, the time as measured at the frame S.

Now, we can proceed in obtaining the variations of the wave properties in a moving frame of reference with respect to those in an ARF. By simplicity we shall use in this section and the next one only plane waves, which are valid a good approximation to the actual spherical waves in regions of space far enough from the wave sources.

Therefore, we can write the plane wave solutions for the wave equation (2) as

$$\mathbf{E} = \mathbf{E}_0 \sin \Phi_0, \qquad (3)$$

where $\Phi_0$ is the absolute phase of the plane wave:

$$\Phi_0 = \mathbf{k}_0 \cdot \mathbf{r} - \omega_0 t. \qquad (4)$$

Here is the constant vector whose magnitude gives the amplitude of oscillation of the field $\mathbf{E}$ and is perpendicular to the corresponding magnetic field $\mathbf{B}$, which has similar equations as (2) and (3) and will not be shown here for simplicity. Further it is well known that $\mathbf{E}$ and $\mathbf{B}$ are perpendicular to the direction $\mathbf{k}_0$ of propagation of the wave and that $\omega_0$ is the constant phase frequency of the wave. These parameters are related by the relation

$$k_0 = \frac{\omega_0}{c}, \qquad (5)$$

which is obtained by calculating the time derivative of the constant phase condition $\Phi_0 = const$.

The direction of $\mathbf{k}_0$ is given by the direction-cosines:

$$\mathbf{k}_0 = k_0 (\cos\phi, \cos\eta, \cos\zeta), \qquad (6)$$

where $(\phi,\eta,\zeta)$ are the angles that $\mathbf{k}_0$ makes with the (x,y,z) axis respectively. Since they undergo the trigonometric identity

$$\cos^2\phi + \cos^2\eta + \cos^2\zeta = 1, \qquad (7)$$

the amplitude of $\mathbf{k}_0$ is $k_0$. Through this, the phase (4) may be rewritten as

$$\Phi_0 = \frac{\omega_0}{c}(x\cos\phi + y\cos\eta + z\cos\zeta - ct). \qquad (8)$$

Finally, the wave length $\lambda_0$ in the frame S undergoes the relation

$$\lambda_0 = \frac{2\pi}{k_0}, \qquad \lambda_0 \omega_0 = 2\pi c, \qquad (9a,9b)$$

which were derived from (3) by observing that the wave should be periodic both in space and time in order that the field $\mathbf{E}$ can be a continuous function of space and time coordinates.
At this point it can be asked what would be the effect of a uniform translation of coordinates in the direction of positive *x* and with constant velocity *v* on the properties of the plane wave (3). This is performed by undergoing (1) and (2) to a Galilean transformation

$$\xi = x - vt,  \tag{10}$$

with y,z remaining unchanged. Here $\xi$ is the transformed *x* coordinate in the new frame of reference S$^*$, which moves parallel to the *x* axis of S with constant velocity *v*, the other coordinates, including the absolute time t, remaining unchanged. Thus (1) becomes

$$\nabla^* \times \mathbf{E} = -\frac{1}{c}\left(\frac{\partial \mathbf{B}}{\partial t} - v\frac{\partial \mathbf{B}}{\partial \xi}\right), \qquad \nabla^* \cdot \mathbf{E} = 0, \tag{11a,11b}$$

$$\nabla^* \times \mathbf{B} = \frac{1}{c}\left(\frac{\partial \mathbf{E}}{\partial t} - v\frac{\partial \mathbf{E}}{\partial \xi}\right), \qquad \nabla^* \cdot \mathbf{B} = 0, \tag{11c,11d}$$

where the star superscript means partial derivation with respect to the variables $\xi$,y,z.

From the new form of the ME (11) we can easily obtain the new form for the electric field wave equation

$$\frac{1}{\beta^2}\frac{\partial^2 \mathbf{E}}{\partial \xi^2} + \frac{\partial^2 \mathbf{E}}{\partial x^2} + \frac{\partial^2 \mathbf{E}}{\partial z^2} = \frac{1}{c^2}\left(\frac{\partial^2 \mathbf{E}}{\partial t^2} - 2v\frac{\partial^2 \mathbf{E}}{\partial \xi \partial t}\right), \tag{12}$$

which can not be written in a simple form as (2) due to the mixed second order derivative and the contraction factor

$$\beta^{-1} = \sqrt{1 - \frac{v^2}{c^2}}. \tag{13}$$

Although (12) has a more complicated form than (2), it will become evident that it shares all the mathematical properties of the later, except the apparent velocity of propagation, which will be named here c', as measured in S$^*$, whose determination is made as follows. The solution of (12) is of the same form as (3):

$$\mathbf{E} = \mathbf{E}_0 \sin \Phi, \tag{14a}$$

$$\Phi = \mathbf{k} \cdot \mathbf{r}^* - \omega t, \tag{14b}$$

where $\mathbf{r}^* = (\xi, y, z)$, which is position vector in S$^*$ and $\mathbf{E}_0$ is the same that appears in (3). Now, the new wave vector is $\mathbf{k} = (k_\xi, k_y, k_z)$, which is related with the other parameters as

$$k = \frac{\omega}{c'}. \tag{15}$$

We can also get a relationship between the wave properties in S and S$^*$ through (10) as

$$\Phi = \xi k_\xi + y k_y + z k_z - \omega t, \tag{16a}$$

$$= x k + y k_y + z k_z - (\omega + v k_\xi) t. \tag{16b}$$

By observing now that the phase $\Phi$ is a scalar quantity, it must have the same value in all coordinate systems, namely, the phase should remain invariant when we pass from a coordinate system to another one. Therefore, equating $\Phi = \Phi_0$, with the later rewritten as

$$\Phi_0 = x k_{0x} + y k_{0y} + z k_{0z} - \omega_0 t. \tag{17}$$

Thus, by comparing (16) and (17), we conclude that the wave vectors old and new should have the same magnitude and direction:

$$\mathbf{k} = \mathbf{k}_0. \tag{18}$$

This means that the light wave does not suffer any change in direction in the new frame with respect to the old.

However, oscillation frequency in the new frame is no longer the same as in the old one, because the motion of the observer produces a change in the apparent rate at which the wave fronts are counted in his frame with respect to the emission rate at the ARF. Therefore, equating the time terms in (16) and (17) yields

$$\omega = \omega_0 \left(1 - \tfrac{v}{c}\cos\phi\right), \tag{19a}$$

where use hás been made of the relationships

$$k_\xi = k_{0x} = \frac{\omega_0}{c}\cos\phi, \tag{19b}$$

which were obtained from (5),(6) and (18).

Of course (18) implies that the wave length remains the same in the new frame, since the source does not move *in fact*, but only apparently, because the observer takes his own frame as being at rest, so that

$$\lambda = \lambda_0, \tag{20}$$

Finally, the apparent velocity c' of the wave in the new frame can be determined from (15), by using (5), (18) and (19):

$$c' = c - v\cos\phi, \tag{21}$$

which can be slightly greater, equal or slightly less the velocity of light c, depending upon the direction of motion of the observer with respect to that of the light wave. From (21) we can immediately interpret (19) as giving the ratio between the apparent speed of light in S' to the absolute speed of light in S, times the absolute frequency

$$\omega = \omega_0 \frac{c'}{c}. \tag{19'}$$

The relation (21) might also be obtained directly from the wave equation (12), by using (14) and (18), that is,

$$k_\xi = k_0 \cos\phi, \quad k_y = k_0 \cos\eta, \quad k_z = k_0 \cos\zeta, \tag{18'}$$

and by using again (15) together with (7), yields the quadratic equation in c':

$$c'^2 + 2vc'\cos\phi - \left(c^2 - v^2\cos^2\phi\right) = 0. \tag{22}$$

Now, equation (22) possesses two solutions, one for a regressive wave, for which $c' = -(c + v\cos\phi)$, which will not be used here, and the other for a progressive wave, given just by (21), which, by hypothesis, is being the only considered here.

At this point, it should be immediately pointed out that (21) *does not come from the simple vector composition* $c' = |\mathbf{c} - \mathbf{v}|$ *between the velocity of the observer and the velocity of light*, as believed in the late XIX century by Lorentz and others. In fact, the vector composition coincides with (21) *if and only* if $\cos\phi = \pm 1$, that is, if the light wave and the observer are moving in parallel or antiparallel to each other. For example, for an angle giving by $\cos\phi = v/c$, which is the absolute angle between the light wave and the perpendicular arm of the Michelson's interferometer (see figs 1 and 2) when the other arm is parallel to the velocity of the Earth, whose magnitude is *v*. Hence, in this case, the apparent velocity of light would be given by the ordinary vector rule by

$$c'_v = c\sqrt{1 - \frac{v^2}{c^2}} = \frac{c}{\beta}. \tag{23a}$$

On the other hand, we obtain from the electromagnetic rule (21), by substituting $\cos\phi = v/c$, the following expression for the apparent velocity of light in the same arm of the interferometer

$$c' = \frac{c}{\beta^2}, \tag{23b}$$

which differs from $c'_v$ above by a root factor $\dfrac{1}{\beta}$, which is just the "contraction" factor that Lorentz (and Fitzgerald) had to introduce *ad hoc* in his analysis of the MMIE in order to account for the "null" result of that experiment within the classical framework. In contrast, this null result arises naturally here.

Thus, the wrong hypothesis of the vector composition between the light and observer velocities was responsible by all the conceptual confusion and misunderstood regarding to the apparently null result of that experiment, which, as it will be explained in details in the section 4 below, remains completely within the classical framework.

However, before we can make this analysis we need firstly to complete the analysis of the wave properties in a moving frame, from which it will be deduced an expression for the apparent velocity c' as a function of the apparent angle $\Phi'$ that the light wave makes with the direction of motion of the Earth, as measured in the moving frame, without which it is not possible to analyze the experiment in this frame. This will be done in the next section.

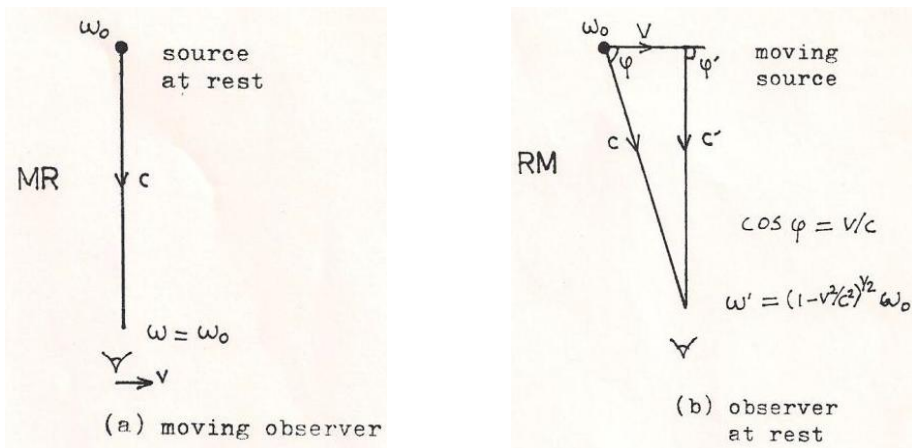

Figure 1. (a) In the MR point of view there is no Transversal Doppler Effect while in (b), in the RM point of view, there is.

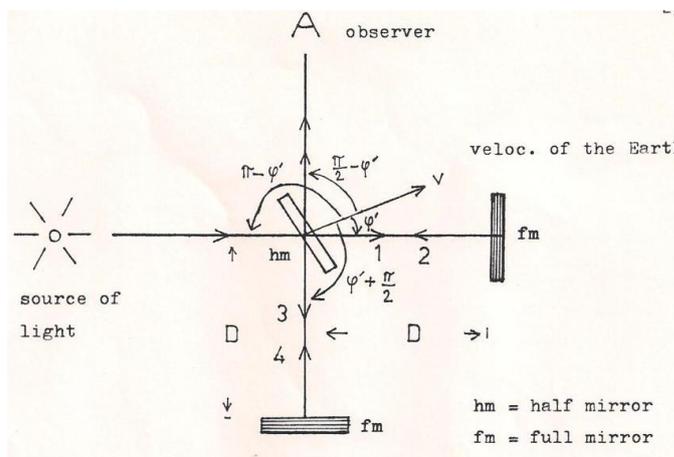

Figure 2. The Michelson's interferometer.

## 3. Transformation of the ME and wave equation under a Voigt's transformation [18].

Now we can make a new transformation of equation (12), which is written in the point of view of a moving observer with respect to a source which stays at absolute rest (MR), to the point of view of an observer which stays at rest relative to a moving source with the same absolute velocity *v* of the observer (MM). This transformation can be easily obtained through a translation together with a dilatation of the absolute time coordinate t, through which is meant to answer the following question: In which moving frame of reference would the phase of a light wave emitted by a source at absolute rest have, at a given point P of absolute space, the same phase as it would have another light wave emitted by an identical source at the moving frame? This question was originally answered, according to Lorentz, by W. Voigt around 1887 [18], in his studies of the properties of the Doppler Effect, as resulting from the form of the solutions of the wave equation. Such a transformation can be written as

$$x' = \beta\xi, \quad y' = y, \quad z' = z, \quad t' = \frac{t}{\beta} - \frac{\beta v \xi}{c^2}. \tag{24}$$

It can be immediately noted that this transformation implies that, in fact, *only the source be in motion*, while the observer stays at rest in his original frame of reference – whether it be at motion or not. This can be seen by applying the transformation to the wave equation (2), which is hence transformed into (12), without the need of supposing a moving observer.

By applying this transformation to the ME in a moving frame (11), the result will be again the original form (2) of them, because this time the source is put in motion with the same velocity of the observer. Thus the observer is now at rest with respect to this source, from what result similar equations as at the absolute frame, but now with all the quantities transformed to the new frame S' because the corresponding source is in fact *in motion*:

$$\nabla' \times \mathbf{E}' = -\tfrac{1}{c}\frac{\partial \mathbf{B}'}{\partial t'}, \qquad \nabla' \cdot \mathbf{E}' = 0, \tag{25a,b}$$

$$\nabla' \times \mathbf{B}' = \tfrac{1}{c}\frac{\partial \mathbf{E}'}{\partial t'}, \qquad \nabla' \cdot \mathbf{B}' = 0. \tag{25c,d}$$

Here the prime signs mean that the differential operations are taken in the new coordinates x',y',z',t' of the new frame of reference and the transformed fields are given by the usual rule of the tensor calculus $A'_\mu = \frac{\partial x'^\mu}{\partial x^\nu} A'_\nu$ [17]. However, their particular forms are of no interest here.

In conclusion, the wave equation (12) in the moving frame MR transforms through the Voigt's transformations (24) – or straightforwardly through the new ME (25) – again to the form of (2), namely

$$\Delta' \mathbf{E}' - \tfrac{1}{c^2}\frac{\partial^2 \mathbf{E}'}{\partial t'^2} = 0, \tag{26}$$

where now the Laplacian operator is taken in the new coordinates x',y',z'.

The new time t' has no metaphysical or subjective character. It is only the time as counted by an electromagnetic clock, whose rhythm varies with the velocity of the electromagnetic source – differently to what happens with a mechanical clock, whose time base is not affected by the velocity of the source – which is used as the base for time counting. However, the objective problem of how to fit this time base in order that the true absolute time t can be measured will be left for a future discussion. For a while we shall continue speaking rather vaguely of the absolute time as the time t.

Now we can relate the wave properties as measured at the frame in motion with respect to a fixed source MR with the same as measured at a fixed frame with respect to an identical source in motion MM. In order to perform this we rewrite the phase $\Phi$ at MR in the form (8):

$$\Phi = \frac{\omega}{c'}(\xi\cos\phi + y\cos\eta + z\cos\zeta - c't), \qquad (27)$$

which can be transformed to MM immediately through the Voigt's transformation (24), when the new phase $\Phi'$ is written as

$$\Phi' = \frac{\omega'}{c}(x'\cos\phi' + y'\cos\eta' + z'\cos\zeta' - ct'). \qquad (28)$$

It should be noted that, since *the source is really in motion*, it will be observed actual modifications in the wave properties, mainly in the wave length and in the phase frequency that are due exclusively to the real motion of the source.

Hence we can relate the wave properties in $S^*$ and S' by transforming (27) in (28) through the inverse transformation of (24), which is not shown here, or we can alternatively go from (28) to (27) by using (24) directly. In any case, from the equality of the phases $\Phi = \Phi'$, since they are invariant, that is, must have the same value in any inertial frame, we get the set

$$\omega' = \beta\omega, \qquad (29)$$

$$\cos\phi' = \frac{c}{c'}\left(\cos\phi - \frac{v}{c}\right), \qquad (30)$$

$$\cos\eta' = \frac{c}{\beta c'}\cos\eta, \qquad (31)$$

$$\cos\zeta' = \frac{c}{\beta c'}\cos\zeta, \qquad (32)$$

where (21) has been used to reach the relations above.

It must be remarked that in the present case the changes in the oscillation rhythm and in the place where the values of the wave phase are the same in the two frames yield, due to the Voigt's transformation (24), an absolute change in the oscillation frequency of the wave, which is not caused merely by the additional counting of the wave fronts by the moving observer, as in (19), but it is in fact due to an actual acceleration of the rhythm of the oscillations, given by (29). There is also a real change in the direction of motion of the wave, given by (30-32), as well as in the magnitude of the wave vector, or equivalently, of the wave length of the wave, which is only possible if the source *is really in motion*.

The magnitude of the new wave vector becomes

$$k' = \frac{\omega'}{c}, \qquad (33)$$

and their components, from the relations (29-33) become

$$k'_x = k'\cos\phi', \quad k'_y = k'\cos\eta', \quad k'_z = k'\cos\zeta'. \qquad (34)$$

Of course they also satisfy

$$k'^2 = k'^2_x + k'^2_y + k'^2_z, \qquad (35)$$

which may be seen with aid of some algebra to be equivalent to equation (22) for c'.

It remains to obtain a new expression for the wave length in the new frame S', because it suffers alterations due to the existence of the absolute motion. Since of course $\lambda' = \frac{2\pi}{k'}$, by using (29) and (33), we get,

$$\lambda' = \frac{2\pi c}{\beta \omega}. \qquad (36)$$

As the wave length in S and S* are the same, given by (8a) and (20), we should used (19') in order to obtain the wave length in the stand point of MM with respect to MR (which is the same as in RR). It follows therefore that

$$\lambda' = \frac{c}{\beta c'} \lambda_0. \qquad (37)$$

Equation (37) tell us that it is possible to detect the motion of a given frame of reference *in its own frame*, because the motion is absolute. This is performed simply by comparing the wave length of a light wave emitted by a light source which stay in its own frame, with that coming from an identical source of light which stays at absolute rest, or that can be considered as such. A possible example would be, within the scope of MMIE, to compare the wave length of a hydrogen light source which stay at the Earth with an equivalent source which stay at the Sun. On the base in the present discussion, it will be expected that the wave length of the source at the Earth should be slightly changed with respect to that of the similar source at the Sun, which would be named here $\lambda_0$, by a quantity given by (37). Neglecting the drift motion of the solar system as a whole – once it produces the same effect in both source and therefore can not be detected – the Sun can be considered as being at absolute rest, so that the major velocity to be considered is that of the translation motion of the Earth around the Sun (neglecting the small effects of other rotation effects that are present in the total velocity of the Earth).

In this situation the relative change in the wave lengths Sun-Earth would be

$$\frac{\Delta \lambda}{\lambda_0} = \frac{v}{c} \cos \phi \leq 10^{-4}, \qquad (38)$$

where we have used a first order approximation in v/c to equation (37) and assumed $\beta = 1$ because (38) is quite small. This implies of course that the detection of the effect is very difficult, but it is not out of the range of measurements of the current spectroscopic techniques. It is possible also that a careful reanalysis of the existing data on Stellar Doppler Effect could be enough to throw some light on this question and uncover the effect [19].

It should be noted that with respect to the MMIE no such an effect as discussed above could be detected, because the light source used in the experiment is the same for the two arms of the interferometer, whether it be of the Sun or it be of the Earth.

## 4. The Lorentz transformation.

It should be noted that it is possible, of course, to pass also from the RR stand point to the RM one, by composing the RR and RM points of view, that is by composing the Galileo's transformation (10), which gives MR, with the Voigt's one (24), which gives MM, that is

$$x' = \beta(x - vt), \quad y' = y, \quad z' = z, \quad t' = \beta\left(t - \frac{vx}{c^2}\right), \qquad (39)$$

which is just the well known form of the Lorentz transformation and which therefore makes possible to relate the wave properties in the frames S and S' directly, without the need of the intermediary frame S*. Therefore this transformation makes possible to pass from the point of view of source and observer at absolute rest to that of source at absolute motion and observer at absolute rest.

However, in his 1905 paper, Einstein obtained the transformation (39) by comparing the points of view RR and MM [8], where the wave equations have the same form, and hence the transit time for a light wave to travel a given distance and then to return to the starting point is apparently the same in these two frames[2], which was the condition imposed by Einstein to get a system of partial differential equations among the coordinates of RR and MM, whose solutions were just the Lorentz transformations (39).

Now, in order to complete the picture of transformations, we can compare the light properties as observed in a frame which moves together with the source MM, with that as measured by an observer at rest together the source at absolute rest RR, from which it will finally result the light properties as seen by an observer at absolute rest with respect to the moving source RM. Further, the wave length is of course given by (37).

On the other hand, in order to obtain a relationship between the frequencies of oscillation of the moving source and of the source at absolute motion, it should be composed the relations (29), which is in a stand point MM, with (19'), written in a stand point MR, from what it will finally result in a stand point RM, namely

$$\omega' = \beta \omega_0 \frac{c'}{c}. \qquad (40)$$

Of course, the relations (40) and (37) are the same that were obtained in the STR, as expected from the previous discussion.

It should be observer that (19') and (40) differ only by the dilatation factor $\beta$, given in (13), which makes the frequency of the moving source, discounted the Doppler effect, greater than the frequency of an identical source which stay at rest. It were not $\beta$, equation (19') in MR and (40) in RM would be exactly the same and it would there be really a complete relativity between these two points of view.

The factor $\beta$ hence breaks the symmetry between the MR and RM situations, making possible to distinguish experimentally between them[3].

A more exact way of making clear the distinction between absolute rest and absolute motion states is the measurement of the so called Transversal Doppler Effect (TDE), which exists clearly only in the situation RM, since in this case the observer receives only the light wave whose direction of propagation makes an absolute angle corresponding to $\cos\phi = v/c$, because only in this case, from (30), the apparent angle of propagation of the light wave would be $\phi' = \pi/2$, namely, the observer would perceive the light wave as if coming from a perpendicular direction with respect to the emitting source. According to (40), by replacing into it the value given above for $\cos\phi$, it would be obtained for the transformed frequency the value

$$\omega' = \frac{\omega_0}{\beta} = \omega_0 \sqrt{1 - \frac{v^2}{c^2}}. \qquad (41)$$

On the other hand, according to (19'), because in the situation MR we would have $\phi' = \pi/2$ only when the very absolute angle has the same value, namely $\phi = \pi/2$. This means that in this case the observer receives a light wave – that he believes to be perpendicular – that really comes from a light wave whose direction of propagation is perpendicular to him, which notwithstanding had left its emitting source in a time early enough to meet the observer in the perpendicular position.

---

[2] In truth it would be greater in MM than in RR, but this could be perceived only out of MM, i.e., in RM.
[3] This is the case of the famous "Twin Paradox" of the STR [8], which in the present view is not a paradox, but simply means that it is possible to distinguish between the states of absolute motion and absolute rest.

Thus, in this case $\cos\phi = 0$ so that (19') implies that

$$\omega = \omega_0. \qquad (42)$$

In conclusion, (42) means to say that it is possible to decide through the TDE whether be the source *or* the observer that is *really* moving, since in each case the Doppler frequencies would be different, as discussed above.

Moreover, a great deal of metaphysical misunderstands happens in the STR by extending the equivalence discussed above among the phases of wave fronts emitted by sources at rest and in motion *to the properties of space and time in themselves* [8], that would make impossible the absolute simultaneity, because the time in itself would become dependent on the position of space where the clock were placed, which seems logically absurd.

Finally, before we can proceed in the reanalysis of the MMIE, we must obtain a relationship expressing the apparent velocity c' in the coordinates of S', that is, as a function of $\phi'$, because the absolute angle $\phi$ can not be measured in a moving frame. This is done by taking the inverse relation of (30), by interchanging the roles of primes and signs, namely

$$c' = \frac{c}{\beta^2 \left(1 + \frac{v}{c}\cos\phi'\right)}. \qquad (43)$$

## 5. Reanalysis of the Michelson-Morley Interference Experiment (MMIE).

Figure 2 shows the Michelson's interferometer when the velocity of the Earth makes an apparent angle $\phi'$ with the horizontal arm of the apparatus. From this configuration is possible thus to determine the absolute difference in the paths followed by the light waves in each of the interferometer arms, since we have already expressed, through (43), the apparent velocity c' with respect to $\phi'$ measured by the moving observer.

The calculation of the apparent velocity of the light waves in each numbered path gives

path1: $$c'(\phi') = \frac{c}{\beta^2 \left(1 + \frac{v}{c}\cos\phi'\right)}, \qquad (44)$$

path2: $$c'(\pi - \phi') = \frac{c}{\beta^2 \left(1 - \frac{v}{c}\cos\phi'\right)}, \qquad (45)$$

path3: $$c'(\phi' + \tfrac{\pi}{2}) = \frac{c}{\beta^2 \left(1 - \frac{v}{c}\sin\phi'\right)}, \qquad (46)$$

path4: $$c'(\tfrac{\pi}{2} - \phi') = \frac{c}{\beta^2 \left(1 + \frac{v}{c}\sin\phi'\right)}. \qquad (47)$$

The corresponding absolute time of transit in the several numbered paths of fig. 2 can now be determined as being

$$t_1 = \frac{D}{c'(\phi')} = \frac{D}{\beta^2 c}\left(1 + \frac{v}{c}\cos\phi'\right), \qquad (48)$$

$$t_2 = \frac{D}{c'(\pi - \phi')} = \frac{D}{\beta^2 c}\left(1 - \frac{v}{c}\cos\phi'\right), \qquad (49)$$

$$t_3 = \frac{D}{c'(\phi' + \frac{\pi}{2})} = \frac{D}{\beta^2 c}\left(1 - \frac{v}{c}\sin\phi'\right), \tag{50}$$

$$t_4 = \frac{D}{c'(\frac{\pi}{2} - \phi')} = \frac{D}{\beta^2 c}\left(1 + \frac{v}{c}\sin\phi'\right). \tag{51}$$

Therefore the total time of transit in the horizontal arm of the interferometer is

$$t_{12} = t_1 + t_2 = \frac{2D}{\beta^2 c}, \tag{52}$$

and the corresponding time of transit in the vertical arm is

$$t_{34} = t_3 + t_4 = \frac{2D}{\beta^2 c}, \tag{53}$$

which are, as can be immediately be noted, equal and independent of $\phi'$. Thus the absolute path difference becomes

$$\Delta D = c\Delta t = c(t_{12} - t_{34}) = 0, \tag{54}$$

which is hence identically null – what confirms a priori the "negative" result of the MMIE.

The conclusion is thus that it is not possible to use the Michelson's interferometer, in the configuration proposed originally, to perform a "positive" detection of the velocity of the observer (of the Earth), through the variation of light properties. This occur because whatever may be the choice for the two paths in which the separated light waves propagate before they reencounter each other, the time of transit will be always the same[4], which leads therefore always to a null total deviation.

Thus, the only practical way to perform a "positive" determination of the observer velocity would be to make use of a non homogeneous optical medium. A possible variation of the Michelson's interferometer for this purpose is shown if fig. 3. The velocity of light in the medium of refraction index $n$ becomes

$$c'_n(\phi') = \frac{c}{n\beta_n^2\left(1 + \frac{nv}{c}\cos\phi'\right)}, \tag{55}$$

where now

$$\beta_n^{-1} = \sqrt{1 - \frac{C^2 v^2}{c^2}}. \tag{56}$$

In this way, the relevant transit times become

$$t_1 = \frac{D}{c'(\phi')} = \frac{\beta^2 D}{c}\left(1 + \frac{v}{c}\cos\phi'\right), \tag{57}$$

$$t_3 = \frac{D}{c'_n(\phi')} = \frac{n\beta_n^2 D}{c}\left(1 + \frac{nv}{c}\cos\phi'\right), \tag{58}$$

where we have assumed $n_{air} \cong 1$ for simplicity.

Finally, the net deviation would become

$$\Delta D = c(t_3 - t_1) \cong (n^2 - 1)\frac{v}{c}D\cos\phi', \tag{59}$$

which shows clearly the "positive" effect that was absent in the original MMIE. $\Delta D$ is of first order in v/c and proportional to D, so that. if it really exists, its detection would be relatively easy to perform. Of course the effect disappears if $n=1$, that is, if the plate $D$ is removed.

---

[4] Since the total geometrical path be optically homogeneous, symmetric and closed.

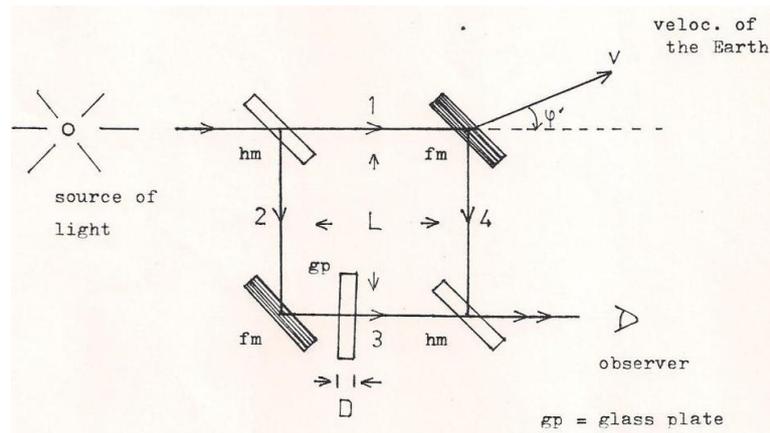
Figure 3. An interferometer that can detect the motion of the Earth.

In conclusion, we should remark that even the original MMIE did not produce an identically null effect, instead of this it was detected a "noise" that varied with $\phi'$ having a very small amplitude, but whose direction of variation agreed with the astronomical motions of the Earth. This was noted by Miller [20], a follower of Michelson, who performed a series of highly elaborated experiments until 1933, from what he obtained deviations still numerically small, although remarkably greater than that of the MMIE and completely in accordance with the Earth's astronomical motions. These deviations are therefore, in the present context, to be considered as purely spurious, that is, as a chaotic effect due to the residual asymmetry of the optical medium used, which included several lenses, mirrors and half glass plates [20]. Thus, the Miller's deviations had amplitudes greater than the Michelson's ones because Miller used a more complex apparatus than the original Michelson's interferometer.

Any way, only the realization of a really "positive" experiment – as that proposed above – could provide a definitive answer to the existence or not of the deviation.

**References.**